\documentclass[11pt, executivepaper]{article}
\usepackage[utf8]{inputenc}
\usepackage[T1]{fontenc}
\usepackage{natbib}
\usepackage{amsmath}
\usepackage{color}
\usepackage{amsfonts}
\usepackage{graphicx}
\usepackage{enumitem}
\usepackage{geometry}
\usepackage{hyperref}
\hypersetup{colorlinks= true, allcolors=blue}
\setcitestyle{aysep={}}

\title{\textbf{Stochasticity and Bell-type Quantum Field Theory}}

\author{Andrea Oldofredi\thanks{Contact Information: University of Lausanne, Dept. of Philosophy,
1015 Lausanne, Switzerland. E-mail: Andrea.Oldofredi@unil.ch}}

\begin{document}

\maketitle 

\begin{abstract}
This paper critically discusses an objection proposed by H. Nikoli\'c against the naturalness of the stochastic dynamics implemented by the Bell-type Quantum Field Theory, an extension of Bohmian Mechanics able to describe the phenomena of particles creation and annihilation. Here I present: (i) Nikoli\'c's ideas for a pilot-wave theory accounting for QFT phenomenology evaluating the robustness of his criticism, (ii) Bell's original proposal for a Bohmian QFT with a particle ontology and (iii) the mentioned Bell-type QFT. I will argue that although Bell's model should be interpreted as a heuristic example showing the possibility to extend Bohm's pilot-wave theory to the domain of QFT, the same judgement does not hold for the Bell-type QFT, which is candidate to be a promising possible alternative proposal to the standard version of quantum field theory. Finally, \emph{contra} Nikoli\'c, I will provide arguments in order to show how a stochastic dynamics is perfectly compatible with a Bohmian quantum theory.
\end{abstract}
\vspace{5mm}
\begin{center}
\emph{Accepted for publication in Synthese}
\end{center}
\clearpage

\tableofcontents
\vspace{5mm}

\section{Introduction: Quantum Field Theory and Primitive Ontology}
In the XX century physicists and philosophers of physics working on the foundations of Quantum Mechanics (QM) have demonstrated that its standard account, albeit extremely empirically successful, is plagued by conceptual difficulties which prevent us from consider it a coherent description of the physical phenomena taking place at the quantum length-scales. 

To overcome these problematic aspects of the theory a significant number of solutions has been presented. Here I will consider the option proposed by the Primitive Ontology (PO) approach, a philosophical perspective which finds its roots in the work done by J. S. in the foundations of non-relativistic quantum physics.\footnote{The main ideas concerning Bell's view on the ontology of physical theories can be found in \cite{Bell:1975aa}.}

The proponents of this perspective have shown that it is possible to solve the conceptual conundrums of QM constructing theories with a clear primitive ontology, i.e. theories which specify what theoretical entities represent real and fundamental objects in the world and how these move in space and time.\footnote{Details concerning this methodology are contained in \cite{Allori:2008aa}, \cite{Allori:2013aa} and \cite{Esfeld:2014ac}.} Following this methodology several theories have been proposed and classified under the label ``Quantum Theories without Observer'' (QTWO); notable examples are the Ghirardi-Rimini-Weber spontaneous collapse theories, in particular the variants GRWm and GRWf which implement a matter density field or flash ontology respectively, Nelsonian mechanics and Bohmian Mechanics (BM). 

This paper is concerned only with the latter theory and in particular with its extensions able to explain the phenomena of particles creation and annihilation typically observed in Quantum Field Theory (QFT); thus, the former proposals will not be considered in the next sections.
\vspace{2mm}

In brief, Bohmian Mechanics is a deterministic quantum theory of particles which move in three-dimensional physical space and follow continuous trajectories. This theory is statistically equivalent to the standard formulation of quantum mechanics although their physical content is remarkably different, since the former makes a precise metaphysical hypothesis concerning the intrinsic corpuscular nature of matter. Hence, in BM\footnote{In this paper I follow the interpretation of BM contained in \cite{Durr:2013aa}, which differs with respect to Bohm's original version of the pilot-wave theory, where the wave function is considered a real physical field.}, every physical fact is reduced to the motion of the Bohmian particles guided by the wave function: according to this theory, physical systems are described by a couple $(\psi, Q)$, where the first element is the usual wave function and the second represent a specific $N$-particle configuration with positions $(Q_1,\dots, Q_N)$. In order to complete the structure of the theory, we need to introduce \emph{two} dynamical laws: on the one hand, the Schr\"odinger equation for the wave function $\psi=\psi(q_1,\dots, q_N, t)$
\begin{equation}
i\hbar\frac{\partial{\psi}}{\partial{t}}=-\sum^{N}_{k=1}\frac{\hbar^2}{2m_k}\Delta_k\psi +V\psi \nonumber
\end{equation}

\noindent{and}, on the other, the guiding equation for the particles' motion:
\begin{equation}
\frac{dQ_k}{dt}=\frac{\hbar}{m_k}\mathrm{Im}\frac{\nabla_k\psi}{\psi}(Q_1,\dots, Q_N)=v_k^{\psi}(Q_1,\dots, Q_N). \nonumber
\end{equation}
 
\noindent{Since} particles moving in physical space have a definite position, BM naturally recovers the notion of trajectory notoriously absent in standard QM.

The empirical equivalence is achieved via \emph{equivariance}: if we assume that at any arbitrary initial time $t_0$ the particle configuration is distributed according to $|\psi_{t_0}|^2$, then it will be so distributed for any later time $t>t_0$, preserving the Born's distribution (see \cite{Durr:2013aa}, Chapter 2, Sec. 7 for the mathematical justification of this statement). 
As already said, the motivations to consider BM as a serious alternative to the standard quantum theory are very well known: not only the notorious measurement problem vanishes, but also its axioms do not contain physically ill-defined notions such as \emph{measurement} and \emph{observer}, present instead in the standard formulation of QM.\footnote{For an introduction to BM and to these foundational issues see \cite{Durr:2013aa}, \cite{Durr:2009fk}, \cite{Bricmont:2016aa}, \cite{Bell:2004aa} and the fundamental papers \cite{Bohm:1952aa} and \cite{Bohm:1952ab}.} All this is in virtue of the clear ontology posed at the basis of the theory. However, this successful approach faces two challenges: 
\begin{enumerate}
   \item to explain the phenomena predicted by QFT;
   \item to find a fully relativistic formulation.
\end{enumerate}
These issues are vividly debated within the Bohmian community, but unfortunately results concerning the second point are still provisional. Nevertheless, it is worth to stress that an \emph{operational} compatibility with special relativity has been achieved, and this is necessary and \emph{sufficient} to claim that the Bohmian QFTs and the standard formulation of QFT are empirically (i.e. statistically) equivalent. Furthermore, many remarkable achievements in order to attain a genuine relativistic version of BM have been obtained by \cite{Dewdney2001}, \cite{Dewdney2004}, \cite{Hiley2010}, \cite{DGNSZ}, \cite{Nikolic:2006aa} and \cite{Nikolic:2013aa}.
Here the issue concerning a relativistic formulation of the BM is left aside, and the following discussion will be focused on a particular class of extensions of Bohmian mechanics to quantum field theory.
\vspace{2mm}

Phenomena typically observed in the context of QFT are the creation and annihilation of particles; however, within this theory, it is hardly the case that we can properly speak about \emph{particles} in the sense of point-sized objects with a precise localization in physical space. Here particles are defined as excitations of quantum fields, which are obtained after the procedure of canonical quantization of a classical field. I rapidly recall it in order to see why the notion of quantum field does not seem to be free of problematical aspects.

In physics the basic idea behind the concept of field is to attribute values of physical quantities to space-time points; thus, they are defined as functions over some regions of space-time. In QFT, the variables of a field become \emph{quantum operators} acting on some Hilbert space. If in standard QM the canonical conjugate variables of position and momentum are promoted to quantum operators imposing the canonical commutation relations, in a field theory one does the same for a field $\phi_{a}(x)$ and its conjugate momentum $\pi^b(x)$, obtaining the following commutation relations:
\begin{equation}
[\phi_a(x),\phi_b(y)]=[\pi^a(x),\pi^b(y)]=0 \nonumber
\end{equation}
\noindent{and}
\begin{equation}
[\phi_a(x),\pi^b(y)]=i\delta^{(3)}(x-y)\delta_a^b, \nonumber
\end{equation}

\noindent{promoting} the classical field to an \emph{operator-valued quantum field}:
\begin{equation}
\phi(x,t)\rightarrow\hat{\phi}(x,t). \nonumber
\end{equation}

This procedure indicates that the basic notion of QFT depends strictly on the identification between operators and physical properties of quantum systems and, as a consequence, the concept of quantum field depends on the notions as measurement and observable. Therefore, the problems arising from this dependence are the same one faces in ordinary QM; hence, one may conclude that even the basic notions of QFT inherit the same ontological problems of standard QM.\footnote{In this regard the reader may refer to \cite{Barrett:2014aa}. In particular, \cite{Durr:2004c} provides a clear analysis of the operator algebra in QM and BM and the unwelcome consequences of the identification between operators as observables.}

Then, in order to achieve a QFT immune from ill-defined concepts, one may follow the strategy known from non-relativistic QM and pursue a research concerning the ontology of QFT in the context of the Primitive Ontology approach, trying to extend BM to the realm of quantum fields. 

Furthermore, looking at the foundations of QFT one notes that the notion of physical state becomes secondary: the central objects are the scattering processes since the principal aim of QFT seems to be the calculation of the amplitudes of scattering events. With the extensions of BM to QFT we assist to a paradigm shift: the notion of physical state recovers anew its centrality. Bohmian QFTs postulate in the first place the primitive ontology of the theory, providing a description of quantum systems in terms of primitive objects moving in physical space according to specific equations of motion, giving back to the theory the shape of a \emph{mechanical} theory. This is a crucial point: from the scattering-oriented approach to QFT, Bohmian QFTs are inverting the current trend through the re-introduction of the familiar notion of ``evolution of physical states''. 

As usual, the main motivation to extend the PO methodology to QFT is that it offers a consistent solution to the conceptual difficulties affecting the standard approach to quantum theories as the measurement problem, the arbitrary division between the quantum and classical regime, the meaning of the quantum formalism, etc.

\cite{Struyve:2010aa} presents an overview of the several extensions of BM to QFT, and among them there are \emph{stochastic} pilot-wave quantum field theories with a particle ontology.\footnote{This paper discusses also Bohmian QFTs implementing a field ontology.} These theories are the focus of my analysis, and I will devote particular attention to discuss their inherent stochasticity.

More precisely, in this paper I critically discuss a well-established idea for which stochasticity should be less compatible with the structure of a Bohmian theory with respect to determinism. In particular, I dispute a claim contained in \cite{Nikolic:2010aa} according to which the structures of the Bell-type QFT presented in \cite{durr_bell_type_2005} are \emph{unnecessary} and \emph{artificial}: unnecessary since the author shows within his theory how to treat the creation and annihilation of the Bohmian particles without adding stochastic elements to a deterministic theory, and artificial because in the Bell-type QFT the underlying deterministic dynamics is broken at random spacetime points \emph{only} in order to account for the particle creation and annihilation events. 

The aim of the paper is to argue that (i) Nikoli\'c claims are well supported neither from a technical point of view, nor from a historical and philosophical perspective, (ii) the Bell-type QFT provides a better explanation of the phenomena of particles creation and annihilation with respect to Nikoli\'c's theory and (iii) that a Bohmian theory can perfectly be stochastic in so far as it is constructed fulfilling the requirements imposed by the PO methodology.
\vspace{2mm}

The structure of the paper is the following: in Section 2 Nikoli\'c's pilot-wave theory of particle creation and destruction is presented and critically discussed, in particular I will argue that this theory implies unwelcome ontological problems which threaten its plausibility. In Section 3 I introduce Bell's proposal for a Bohmian QFT with a particle ontology and his main concerns about a stochastic dynamics. Section 4 deals wth the Bell-type QFT (BTQFT); furthermore, a full discussion of the issue of stochasticity in the context of BM is provided. The last section contains the conclusions.

\section{A pilot-wave theory of particle creation and destruction}

In \cite{Nikolic:2010aa} has been proposed a theory which makes Bohmian mechanics compatible with relativistic QFT via the introduction of a deterministic dynamics able to describe the variations of the particles' number, and consequently accounting for the phenomena of particle creation and annihilation. This theory, thus, preserves determinism also within the domain of quantum field theory, maintaining a structural continuity with the standard Bohmian approach. For this reason, the author claims that his model provides a more natural explanation for these events with respect to the theory presented in \cite{durr_bell_type_2005} (and consequently in \cite{Bell:1986aa} being the former a generalization of the latter) which, instead, introduces a stochastic evolution governing the primitive ontology in order to yield an explanation of the QFT's phenomenology. Hence, Nikoli\'c explicitly claims that these Bohmian QFTs, being based on artificial and unnecessary structures, are \emph{not} the most natural descriptions available in order to represent the variations of particle number observed in QFT.

Thus, in order to evaluate the soundness of this claim, let us consider Nikoli\'c's theory with a simple example. Suppose a QFT state representing an unstable particle which may decay into two new particles in a given interval of time from $t_0$ to $T=t_{0}+\Delta{t_{0}}$: 
\begin{align}
|\Psi\rangle= |1\rangle+|2\rangle \nonumber
\end{align}
\noindent{accordingly,} the first term of the superposition corresponds to the unstable particle, described by a 1-particle state, and the second term describes a couple of particles, the result of the decay process, with a 2-particle state. Since we do not have access to the actual number of particles prior to a measurement process, we do not know whether or not the particle has decayed within $T$. Nonetheless, it is a fact that in nature such superposition is never observed if measurements of particle number take place. The standard formulations of both QM and QFT resolve this issue via the collapse of the wave function induced by a measurement process: the interaction (and successive entanglement) between quantum and measuring systems will lead to the selection of one among the possible outcomes, so that only one of the superposed states is effectively observed. Unfortunately, given the motivations stated in the previous section, it is well known that this solution is problematic under many aspects. 

To avoid them, Nikoli\'c's recasts the above physical situation in Bohmian terms: his theory holds a particle ontology, therefore, it assumes that there are point-sized localized objects moving in physical space. Thus, Nikoli\'c rewrites the above superposition with an explicit reference to the particles involved:
\begin{align}
\label{superposition}
\Psi(x_1, x_2, x_3) = \Psi_1(x_1)+\Psi_2(x_2, x_3), 
\end{align}
\noindent{where} $ \Psi_1(x_1)$ correspond to the 1-particle wave function describing the unstable particle, and $\Psi_2(x_2, x_3)$ correspond to the 2-particle wave function representing the decay products. 

According to Nikoli\'c's interpretation of this superposition, \eqref{superposition} \emph{does not describe a state of ignorance} about the actual number of particle. Rather, being it a 3-particle wave function (see the LHS), it refers to \emph{three} real particles trajectories, thus, to three particles \emph{simultaneously} existing in physical space. 

However, since observations detect always either the unstable particle or the particles resulting as decay products, it is necessary to suppress this superposition taking into account a measuring apparatus and the decoherence processes originated by its interaction with our QFT state, following the usual Bohmian theory of measurement.\footnote{For a detailed treatment of the measurement theory in BM see \cite{Bohm:1952ab} and \cite{Durr:2009fk}, Chapter 9. Since in his paper Nikoli\'c neither proposes a new theory of measurement nor refers to a particular one, I suppose he tacitly assumes the standard Bohmian theory exposed in the mentioned references.}

If we measure the number of particles, the total wave function is the following:
\begin{align}
\Psi(x_1, x_2, x_3, y) = \Psi_1(x_1)E_1(y)+\Psi_2(x_2, x_3)E_2(y), \nonumber
\end{align}
\noindent{here the variable} $y$ represents the particles' configuration of the measuring apparatus, $E_1, E_2$ represent the possible states of the detector. In this simplified case we are assuming that for all practical purposes $E_1(y)\cap{E_2(y)}=0$, meaning that the wave functions $E_1(y), E_2(y)$ do not overlap in configuration space, so that if $y$ takes its value $Y$ in the support\footnote{The support of the wave function is a region in configuration space where it has non-zero values.} $E_2$, then $E_1(Y)=0$ and vice versa. Suppose now to find two particles after the measurement. Since the interaction between system and apparatus caused what is called an effective collapse of the wave function in one of the possible states of the superposition, the wave function that correctly describes the system is $\Psi_2(x_2, x_3)$. Thus, given the initial superposition prior the measurement of the particles' number, where \emph{three} particles existed in physical space, we conclude the observation with only two particles at locations $x_2, x_3$. The third particle has been \emph{destroyed} by the dynamical interaction with the measuring apparatus.

In this regard, Nikoli\'c's asks a peculiar question: what does happen to the particle in $x_1$? According to the theory, its motion is governed by a dynamical law in 4-dimensional space which is the following:
\begin{align}
\label{Nik_guide}
\frac{dX^{\mu}_1}{ds}= \frac{\frac{i}{2}\Psi^*\overleftrightarrow{\partial}^{\mu}_1\Psi}{\Psi^*\Psi}, 
\end{align}

\noindent{$s$} is an auxiliary scalar parameter along the particle trajectory.\footnote{This equation has been derived in \cite{Nikolic:2006aa}. It is important to mention that \eqref{Nik_guide} is equivariant if and only if we have probability distributions on $\mathbb{R}^{4N}$ with density $|\psi|^2$; this dynamical law fails to be equivariant considering probability distributions on $\mathbb{R}^{3N}$ with density $|\psi|^2$ setting all time variables to $t$. Plausibly, Nikoli\'c may reply that his theory of particle creation and destruction is defined only in $\mathbb{R}^{4N}$, circumventing this objection.} In the case of the superposition considered in our example, after the performance of the measurement, the four components of the 4-velocity associated with the particle at position $x_1$ are zero: the effective wave function that correctly describes the system is $\Psi_2(x_2, x_3)$ which, in turn, does not depend on $x_1$, hence the derivatives in \eqref{Nik_guide} vanish, implying that the total velocity associated to the particle in $x_1$ is zero.

It is crucial to underline that this model not only assigns spatial coordinates to the Bohmian particles, but also a temporal one is associated with them, so that their location in spacetime is completely specified. Since the spatial and temporal coordinates are treated on equal footing, the particle in $x_1$ after the interaction with the measurement apparatus does not change its position in spacetime, becoming a point-sized object with neither \emph{spatial} nor \emph{temporal} extension: ``It can be thought of as a point-like particle that exists only at one instant of time $X^0_1$. It lives too short to be detected. Effectively, this particle behaves as if it did not exist at all'' (\cite{Nikolic:2010aa}, p. 1482). This example shows how the superposition considered in our example is resolved and how creation/destruction events are dynamically induced by this theory. 
\vspace{2mm}

The above discussion rests on the assumption that $E_1(y)\cap{E_2(y)}=0$, nonetheless, considering more realistic situations one should relax it and take $E_1(y)\cap{E_2(y)}\approx{0}$, where the overlap in configuration space is negligible but not exactly null. 

Generally, in standard BM, this detail does not make any substantial difference; however, it has a remarkable ontological consequence for Nikoli\'c's theory. Saying that the overlap of the detectors' wave functions is approximately zero implies that the values of the 4-velocity of the particle located at $x_1$ will be only \emph{approximately} zero. Then, considering realistic physical situations, the destroyed particles will have extremely small values for the 4-components of their velocities (but not null values), thus, they are never, strictly speaking, really destroyed. 

As a consequence, Nikoli\'c claims that they form a sea of inert particles whose contribution to the evolution of the particle configuration is negligible. For all practical purposes, one may say that in the previous example the total wave function has effectively collapsed in one of the possible terms of the superposition (in our case in $\Psi_2(x_2, x_3)$), and the other term effectively vanishes. Nevertheless, taking seriously into account the ontological consequence implied by the condition $E_1(y)\cap{E_2(y)}\approx{0}$, one has to accept that the particle at $x_1$ is not literally destroyed. 

Thus, although from a operational perspective the particle at $x_1$ behaves as if it does not exist, given that its 4-velocity is approximately zero (and this is totally reasonable from a practical perspective), looking at the theory's ontology we must say that it still, somehow, exists. Nevertheless, existence should be considered a ``0/1''-property, meaning that an object either exists or it does not, implying that there are no degrees of existence. Alternatively stated, existence is not a fuzzy property. Hence, if $x_1$ 4-velocity approximates zero, this particle should not be considered really destroyed.  

According to this theory, it is the dynamical interaction between systems and environment what induces destruction and creation events; in turn, this implies that our world abounds of inert particles which cease to contribute dynamically to the evolution of the particles' configuration, but still continue to exist. Therefore, in this theoretical framework there is a number of particles with continuous trajectories in physical space and a number of inert objects, which create a sea of ``dead'' particles, as Nikoli\'c labels them. As already stressed, these latter ones are suppose to live (which means to be detectable) for an infinitesimally short time, so that these dead particles are \emph{practically} not observable. 

After this brief introduction to the mechanisms of particles creation and destruction in Nikoli\'c's theory, let us discuss its implications. 

This theory has a number of notable merits since its maintains a structural continuity with the standard BM being also able to describe the variations of the particles' number accounting for the phenomena of particles creation and annihilation, it is formulated in a relativistic framework and it takes into account the decoherence formalism in the interaction processes between quantum and classical systems. However, despite these positive features, there are significant consequences implied by this theory which threaten its plausibility. 

To explain this point it is necessary to recall the very same QFT state we discussed at the beginning of this section. Suppose the following, highly idealized, situation: the unstable particle is contained in an empty box, without any possibility for it to interact with the surrounding environment, and vice versa (it is not possible to interact directly with the quantum system inside the box). We know that it may decay into two new particles or it may not, according to the superposition $|\Psi\rangle= |1\rangle+|2\rangle$. Before the interaction between system and apparatus, the author claims that \emph{all three particles} associated with wave functions $\Psi_1(x_1)$ and $\Psi_2(x_2, x_3)$ in the superposition \eqref{superposition} do have trajectories in physical space. Thus, before the observation of the particles' number, three particles exist. After the measurement, the total wave function will take value in only one of the possible supports. Suppose that the unstable particle has not decayed, then the 4-velocities $\frac{dX^{\mu}_2}{ds}$ and $\frac{dX^{\mu}_3}{ds}$ \emph{become} zero and we observe just one particle, otherwise, as already said, if the decay process occurs we will find two particles, implying that the velocity of the particle in $x_1$ will vanish.

This simple thought experiment is useful to understand where Nikoli\'c's theory is problematic. Considering a situation in which it is not possible to interact with the quantum system, supposing e.g. to leave the box closed, Nikoli\'c's theory implies that three particles are moving in it. This fact, however, is implausible given the initial conditions of the thought experiment \emph{and} the standard Bohmian theory of measurement, which is assumed by the author; it is a physical impossibility to claim that three particles are simultaneously moving within the box since for every instant of time $t$ within the box there are always \emph{either} one \emph{or} two particles: the superposition of these states refers only to our ignorance about the system's evolution. Thus, there are only two mutually excluding physical possibilities: (i) the case in which the decay will occur, where there will be a precise time $t_d$ in which the unstable particle will decay into the new ones, or (ii) the case in which the decay will not occur, and only the unstable particle will continue to move within the box. 

Therefore, the interaction with a measurement apparatus is essential to \emph{know} in which state the system effectively is, but it is not essential to determinate the state of the particle(s). 
Furthermore, generalizing the conclusion of this thought experiment, one may claim that this theory does not provide a clear description of the particles' behavior when measurements of the particle number are \emph{not} performed: if it is not clear what happens to the particles in situations in which no measurements are performed, it follows that it is also not clear whether this theory is able to provide a clear picture of our reality, which is one of the main motivation to propose a Bohmian theory.

Moreover, it is worth noting that the superposition \eqref{superposition} may generate paradoxes since considering a reference frame in which $X_1$ lies within $X_2$ past light-cone this theory may imply backwards causation, which would certainly be a motivation to discard this theory.

Another example which sheds light on shaky features of Nikoli\'c's theory is the following. Suppose we perform a measurement of the particle number starting from the state \eqref{superposition} obtaining ``2'' at some initial time $t_0$, so that the system is described by a 2-particle wave function, then we do not interact with it, letting it evolve without perturbations. Suppose further that after a time interval $T=t_0+\Delta{t_0}$, the unitary evolution of the system will lead again to a non-trivial superposition of a 1-particle and a 2-particle quantum state. Then we make another measurement of particle number supposing to observe, this time, \emph{one} particle. Thus $X_1^0(s)$ needs to be brought from $t_0$ to $\Delta{t_0}$, resulting in a world line that effectively \emph{exists} throughout this time interval $T$. 

This scenario undermines Nikoli\'c's theory since according to it the world-line of particle in $x_1$ must not exist from time $t_0$ onwards. If the main point of this theory is to claim that although the dynamical equations govern \emph{three} particles moving in spacetime every constant-$t$ hyperplane is intersected \emph{either} by one \emph{or} by two world-lines, then this example shows the opposite. 

Finally, another problematical aspect of this theory regards its ontology, or more precisely, the sea of dead particles. Albeit the idea of having initial and final spacetime points in which particles' trajectories begin and end is elegantly formulated using a multi-time formalism, taking seriously Nikoli\'c's theory, it seems that the destroyed particles are not literally annihilated; thus, even though they are not experimentally observable, still exist and form a sea of dead particles. Given the very large number of particles present in our universe and the number of interactions which take place among them, we should expect a remarkable number of dead particles as well. However, even assuming that the theory somehow is able to provide an operational explanation of the phenomena of particle creation and destruction, it does not behaves equally well from an ontological point of view, since the particles are simply not really destroyed. But, as said above, existence is a property which does not admit a ``continuous'' spectrum of possibilities, namely either an object exist or it does not. Thus, the theory implies what we may call an ontological \emph{surplus} formed by these dead particles, as residual of partial particle destruction processes. These objects, however, are very peculiar ones since it is not at all clear whether or not fulfill the necessary conditions imposed by the particle notion. Are they localized objects? Do they have mass and charge? etc. The answer to this set of questions is only ``approximately no'', which leave us with a sea of very peculiar entities.

In conclusion, even though it is true that Nikoli\'c's QFT does not include any stochastic element in a Bohmian framework, from the arguments given above one may claim that his theory implies a series of unwelcome technical and ontological consequences; thus, it seems appropriate to claim that one should consider different proposal for a successful explanation of QFT phenomenology. 

\section{Bell on Quantum Field Theory}

In \cite{Bell:1986aa} is proposed a model with a particle ontology which reproduces the statistics of any regularized QFT (i.e. any QFT with cut-offs). Providing an empirically equivalent formulation of quantum field theory in Bohmian terms, Bell explicitly showed that an extension of the causal approach to the context where phenomena of particles annihilation and creation are observed is not only plausible and desirable, but also \emph{possible}.

According to this model, the physical 3-dimensional space is replaced by a discrete lattice $\Lambda$, and the local beables for this quantum field theory are the fermion number at each lattice point. To this end, Bell claimed that the \emph{minimal} (but not unique) ontology able to reproduce the experimental evidence supporting QFT is, once again, a particle ontology:
\begin{quote}
What is essential is to be able to define the position of things, including the positions of instruments pointers or (the modern equivalent) of ink on computer output. [...] The distribution of fermion number in the world certainly includes the positions of instruments, instrument pointers, ink on paper and much much more. (\cite{Bell:2004aa}, p.175)
\end{quote}

Thus, a generic configuration is specified by the number of fermions\footnote{Within this theory bosons are not part of the ontology.} $q(x)$ at each lattice site $x\in\Lambda$. Thus, the configuration space of this model is 
\begin{align}
\mathcal{Q}=\Gamma(\Lambda):=\Big\{q\in\mathbb{N}^{\Lambda}:\sum_{x\in\Lambda}q(x)<\infty\Big\} \nonumber
\end{align}
which represents the space of all the possible configurations of a variable but finite number of particles on this lattice (see \cite{Tumulka:2005} for details, here I follow their notation). 

The ontology of this model is slightly different with respect to the standard Bohmian theory since particles' positions do not possess the beable status, and consequently, Bell's model does not provide the trajectories for the fermions moving in space.\footnote{It is important to note that Bell himself repeatedly stressed that the choice of the beables for a given theory is not unique.} Nonetheless, the model is still a particle theory because the lattice fermion numbers are associate with definite particles' positions in space, even though the dynamics of the theory does not describe the motion of single fermions. 

According to this theory, a physical state is fully characterized by a couple where the first element is the fermion number configuration, and the second is the wave function of the system, as in standard BM. 

Another relevant novelty brought about by this model concerns the dynamics, which is stochastic. In order to describe the events of particles' creation and annihilation, Bell provided an equation of motion for the fermion numbers in terms of the jump rate $\sigma_t(q, q')$, which represents the probability for a given configuration of fermions $q$ to jump into another configuration $q'$ with a different number of fermions within a certain interval of time. These equivariant random jumps between two configurations $q$ and $q'$ correspond to the creation and annihilation of particles and assume of the form
\begin{align}
\sigma_t(q, q')=\frac{\frac{2}{\hbar}[\mathrm{Im}\langle\Psi_t|P(q){H}P(q')\Psi_t\rangle]^+}{\langle\Psi_t|P(q')\Psi_t\rangle}. \nonumber
\end{align}
\noindent{where} the sign ``+'' says that these probabilities are non-negative.\footnote{More precisely, the notation $[\dots]^+$  considers only the positive part of the quantity between the squared brackets, setting the value equal to $0$ whenever this quantity is negative; for details see \cite{Tumulka:2005}. Furthermore, the authors show this is a special case of \eqref{bell}, which is the jump rate defined in \cite{durr_bell_type_2005}.} 

The choice to introduce stochastic elements into the dynamics of the theory is motivated by the phenomenology of QFT, which includes literal and random events of particles creation and annihilation. 

Nonetheless, it is appropriate to underline that Bell viewed his theory only as a phenomenological model without any pretension to be rigorously interpreted. There are several motivations to support this claim and agree with Bell: in the first place, this ontology implies discreteness of space, but there is no evidence whatsoever that in QFT space should be treated as a discrete substance. Thus, it seems legitimate to say that this is a very unwanted feature of the theory, and maybe an indication to find a more suitable ontology for QFT. Secondly, the author stated that the stochasticity is unwelcome since 
\begin{quote}
the reversibility of the Schr\"odinger equation strongly suggests that quantum mechanics is not fundamentally stochastic in nature. However I suspect that the stochastic element introduced here goes away in some sense in the continuum limit (\cite{Bell:2004aa}, p. 177).  
\end{quote}

In this regard, from the above quotation it is also plausible to think that Bell viewed stochasticity somehow directly connected to the discreteness of physical space, since he suspected that in the continuum limit this dynamics would have been replaced by a deterministic evolution in a continuum space. 

Finally, Bell thought that the cogent issue to be solved was the achievement of Lorentz invariance, being the model only operationally equivalent with respect to a relativistic quantum theory (it should be recalled here that also the standard regularized QFT is not, strictly speaking, Poincare covariant). Unfortunately, nowadays, all Bohmian models for QFT are only operationally compatible with relativity.\footnote{In this regard it is important to underline different attitudes concerning Lorentz invariance in the context of Bohmian mechanics. One the one hand, there are exponents of the pilot-wave community e.g. Bohm, Valentini and Holland (among others) who think that a detailed microscopic description of the quantum mechanical regime must violate Lorentz invariance, but since it is not possible to have access this description, there will occur no violation of the special theory of relativity; on the other hand, D\"urr, Goldstein and Zangh\`i claimed that a genuine Lorentz invariant Bohmian theory is not in principle excluded. In both cases, however, the notion of absolute simultaneity must be recovered in order to define a guidance equation for the Bohmian particles, violating the spirit of special relativity (see \cite{Lienert:2011aa}, sec. 4.1). For a detailed discussion the reader may refer to the following papers: \cite{DGNSZ}, \cite{Durr:1992aa},\cite{Valentini:1991aa} and \cite{Butterfield:2007aa}, sec. 7.1.}

These facts, I think, are sufficient to claim that this model should not be interpreted literally, in perfect agreement with Bell's ideas. 

Be that as it may, even though within the literature concerning the Bohmian formulations of QFT it has been shown by \cite{Colin:2003aa} how to generalize Bell's model with a deterministic dynamics, it must be admitted that in several places Bohm himself argued that to restore determinism is \emph{not} the main aim of his approach to quantum physics. Remarkable examples can be found (i) in the ninth section of his \cite{Bohm:1952aa}, or (ii) in \cite{Bohm:1954aa}. In the latter paper the authors developed a hidden variable theory with a stochastic background field, instead in the former Bohm discusses two possible manners to modify his theory in order to conceive possible experiments able to distinguish it from ordinary quantum theory (especially at distances of the order of $10^{-13}$ cm or less): the second option consists in a modification of the Schr\"odinger equation, which becomes a stochastic equation. 

Although Bell interpreted stochasticity just a contingent fact of his model, and not as a fundamental feature of nature, it is fair to claim that Bohm thought that at the fundamental level physics may be not deterministic. On this basis, a generalization of Bell's model where the stochastic dynamics is taken seriously has been advanced by \cite{Durr:2004aa} and \cite{durr_bell_type_2005}, for this reason these authors called their model \emph{Bell-type QFT} (BTQFT). As we will see in the next section, this theory provides a different ontology which allows one to interpret it not just as a heuristic framework to show that an extension of BM to QFT is feasible, but as a proper candidate to be a possible alternative to the standard formulation of QFT.

\section{Bell-type Quantum Field Theory}

The Bell-type QFT is a theory of particles moving in 3-dimensional physical space, whose trajectories can randomly \emph{begin} and \emph{end} at certain space-time points. With respect to Bohmian mechanics, this is the main conceptual innovation: the creation events correspond to the beginning of a particle's trajectory, similarly, annihilation events correspond to its end. According to this theory these phenomena are literally interpreted, therefore, it provides a ontology of particles that can stochastically come into existence as well as cease to exist. This is the element that Nikoli\'c contests as artificial. 

Contrary to Bell's model, in BTQFT the beables are again the particles' positions for both particles and anti-particles, consequently there is no need to postulate discreteness of space, avoiding the unnatural feature of Bell's theory. The third important ontological difference with respect not only to this latter model, but also with the usual treatment of bosons in QFT is that, according BTQFT, bosons assume a particle status exactly as fermions. 

As usual in the context of BM, a physical system is described by a pair $(Q_t, \Psi_t)$, where the former correspond to a specific $N$-particle configuration specifying the number and positions of the particles in 3-dimensional physical space, and the latter is the state vector in the appropriate Fock space (i.e. obeying either to the Fermi or to the Bose statistics), and can be seen as a function on a configuration space of a \emph{variable} number of identical\footnote{For details concerning identical particles in BM the reader should refer to \cite{Goldstein:2005b} and \cite{Goldstein:2005a}. For the purpose of the paper it is sufficient to say that, being the particles identical, they are invariant under permutations: instead of having a given configuration of labeled particles, where position 1 is occupied by particle 1 and position $n$ occupied by particle $n$, here we have a set of positions occupied by particles which could be permuted without affecting or modifying the particles' configuration.} particles (see \cite{durr_bell_type_2005} for details on such space). 

In the case one considers only a single species of particles, e.g. the electrons, the configuration space is 
\begin{equation}
\mathcal{Q}=\bigcup_{N=0}^\infty\mathcal{Q}^{[N]}_{e^-}, \nonumber
\end{equation}  
\noindent{where} $\mathcal{Q}^{[N]}=\mathbb{R}^{3N}/permutations$. Taking into account more than a single particle species, the configuration space is the Cartesian product of the involved particle sectors. The example proposed by the authors is the configuration space of Quantum Electro-Dynamics (QED), which involves three different species of particles: electrons ($e^-$), positions ($e^+$) and photons ($\gamma$). In this case we have a Cartesian product of three different configuration spaces: $\mathcal{Q}^{[N]}_{e^-}, \mathcal{Q}^{[N]}_{e^+}, \mathcal{Q}^{[N]}_{\gamma}$, and the total configuration space is given by
\begin{equation}
\mathcal{Q}_{QED}=\bigcup_{N=0}^\infty\mathcal{Q}^{[N]}_{e^-}\times\bigcup_{N=0}^\infty\mathcal{Q}^{[N]}_{e^+}\times\bigcup_{N=0}^\infty\mathcal{Q}^{[N]}_{\gamma}, \nonumber
\end{equation}  
\noindent{providing} information about the particles' number and positions.\footnote{According to the theory, $\Psi_t$ has the habitual double role: on the one hand, it guides the particles' motion, on the other determines the statistical distribution of the particles' positions.}
\vspace{2mm}

In order to check whether or not Nikoli\'c's objections are justified, let us have a closer look to the dynamics of the theory. 

The Bell-type QFT introduces discontinuities in the particles' trajectories in order to take into account the events of particles annihilation and creation. The picture a) below\footnote{This picture is taken from \cite{Durr:2004aa}.} represents the emission of a photon at time $t_1$ (dashed line) from an electron (solid line) and its absorption at time $t_2$ by a second electron. These two events correspond to a creation and annihilation event respectively. Between them the photon evolves according to a deterministic trajectory exactly as the electrons. The picture b) represents a creation of an electron-positron pair at the end of a photon trajectory. 
\vspace{2mm}
\begin{center}
\includegraphics[scale=0.4]{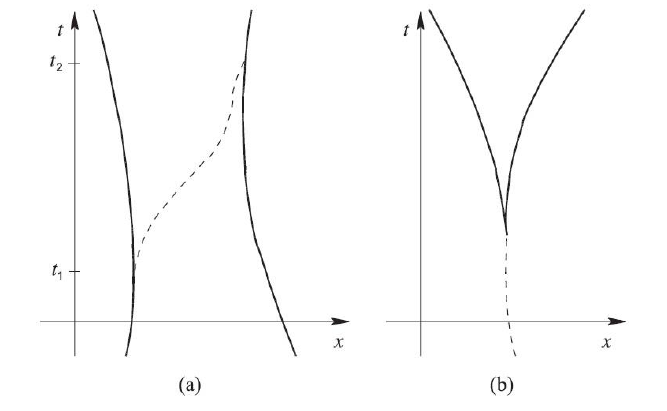}
\end{center}

In these examples we can explicitly see that the number of the particles is not constant: for instance, in the first picture, for every time $t<t_1$ the configuration is composed by only two electrons, then at time $t_1$, a photon is created and the particle number increases: between $t_1$ and $t_2$ the configuration counts three particles. Similarly, at time $t_2$ the photon gets absorbed by another electron and the number of particles decreases. These variations of the particles are to be considered literal events in physical space and imply modifications in the configuration space: at the creation event the particle configuration $Q_t$ jumps in an higher sector, vice versa at each annihilation event it jumps to a lower one. Between the jumps the configuration moves continuously in one sector. 

It is correct to say that the overall dynamics of the theory is a piecewise deterministic Markov process. The jumps introduced here are motivated, as in Bell's case, by the phenomenology of QFT: since we have important experimental evidence speaking in favor of literal particles creation and annihilations events, the authors decided to take it seriously and to propose a dynamics including birth-and-death processes in order to represent them. 
\vspace{2mm}

As usual in BM, this theory provides a set of two equations for the total dynamics. The state vector evolves according to the Schr\"odinger equation:
\begin{equation}
i\hbar\frac{d\Psi_t}{dt}=H\Psi_t \nonumber.
\end{equation}

\noindent{where $H$ is the Hamiltonian operator.} In quantum field theory the Hamiltonian is defined as a sum of terms: $H_{tot}=H_0+H_{Int}$, where the first term correspond to free, non-interacting, processes and the second term describes interactions. 

Accordingly in BTQFT, the free part of the Hamiltonian  $H_0$ defines a velocity field on configuration space, and correspond to the deterministic part of the dynamical Markov process mentioned above. Thus, between the creation and annihilation events the particles follow continuous trajectories obeying to the Bohmian law
\begin{equation}
\label{guide}
\frac{dQ_t}{dt}=v^{\Psi_t}(Q_t). 
\end{equation}

Similarly, the interaction Hamiltonian $H_{Int}$ is associated with the discontinuities of the particles trajectories and correspond to the stochastic part of the overall dynamical process. These jumps are defined by the \emph{jump rates} $\sigma=\sigma(q',q,t)=\sigma^{\Psi_t}(q',q)$, and they correspond to a transition from a given configuration of particles $q$ to another one $q'$ which differs in the number of particles.
The jump rate for $H_{Int}$ is given by
\begin{equation}
\label{bell}
\sigma_t(dq|q')=\frac{[\frac{2}{\hbar}\mathrm{Im}\langle\Psi_t|P(dq)H_{Int}P(dq')\Psi_t\rangle]^+}{\langle\Psi_t|P(dq')\Psi_t\rangle}.
\end{equation}

Considering the picture a) one can easily see that the event of the photon emission correspond to a jump of rate $\sigma_t(q',q)$ where the starting configuration is composed by two electrons and the arrival configuration counts also the photon. At time $t_2$ another jump occurs: in this case the initial 3-particle configuration $q'$ jumps into the final configuration counting two electrons (the configuration may not be equal to the original $q$ since the positions of the two electron have changed). These rates give the probability for a given configuration $q$ at an arbitrary time $t$ to jump in a given interval of time $(t, t+dt)$ into another configuration $q'$. Destinations and times of the jumps are the stochastic elements of the model, and being these jumps Markov processes (i.e. memoryless) they do not depend on the \emph{past} histories of the particles, but depend only on the present state of the configuration and the wave function. 

The total dynamics is, thus, a sum of processes: on the one hand there is the deterministic process associated with the continuous path between the jumps, and on the other there is a stochastic process associated with the particle creation and annihilation events. But the total dynamics forms a unique and coherent process. 

Given the definition of the Hamiltonian operator in QFT, it is more than plausible the idea to associate different processes to the summands of $H$; but it would be a misrepresentation of the theory to claim that this new dynamics is only a deterministic motion \emph{plus} a stochastic element, artificially inserted to represent the variations of the particles' number. Alternatively stated, one has to consider the total dynamical process offered by the BTQFT, which provides a \emph{unique} process for the evolution of the particles' configuration: the deterministic and stochastic processes have to be considered on an equal footing, given that BTQFT incorporates phenomena which standard BM cannot account for. 

Furthermore, although there are many possible choices in order to describe the jump rates, the authors provide a rigorous proof concerning the existence of a unique ``minimal'' jump rate, in the sense that at any time $t$ only one between these two jumps $q_1\rightarrow{q_2}$ and $q_2\rightarrow{q_1}$ is allowed. To complete the picture it should be said that the transitions permitted can only be made to $n+1$ or $n-1$ particles, where $n$ is the number of particles of a generic configuration $q$: a particle can appear, so that the transition to $n+1$-states correspond to a birth processes, and disappear, meaning that the transition to $n-1$-states correspond to death processes (or a particle can be replaced by a pair (pair creation) and vice versa (pair annihilation)).  The results on the global existence, coherence and uniqueness of the jump rates in BTQFT can be found in \cite{durr_bell_type_2005} and \cite{Tumulka:2005}. 

The last step remained to discuss is the empirical adequacy of the Bell-type QFT. In \cite{durr_bell_type_2005} it has been extensively shown that if the particle configuration $Q(t_0)$ is chosen randomly with distribution $|\Psi(t_0)|^2$, then at any later time $t$ it is distributed with density $|\Psi(t)|^2$. This result is the extension of equivariance in the context of QFT.

Since both the free and the interacting part of the Hamiltonian are by construction associated with equivariant Markov processes, equivariance is carried over intact in this extension of BM. Thus, the empirical equivalence has been achieved with the standard formulation of regularized QFT. Equivariance is, as \cite{durr_bell_type_2005} characterize it, ``an expression of the compatibility between the Schr\"odinger evolution for the wave function and the law [...] governing the motion of the actual configuration'', thus, the Markov transition probabilities of this theory, being derived directly from the Schr\"odinger equation, are defined by equivariant generator operators acting on configuration space (see \cite{durr_bell_type_2005}, Sec. 2), which means that the theory is inherently compatible with the structure of QFT. 

The notions of Equivariance and ``process additivity'' are the key features of BTQFT, since they are the guiding principles in the construction of the dynamics: the processes associated with $H_0$ and $H_{Int}$ are defined in a manner which allows to yields \emph{typical} histories for the primitive variables compatible with quantum statistics. Thus, it follows that BTQFT is the natural process associated with $H$ in QFT: the sum of equivariant generators for the transition probabilities defines a unique equivariant process associated with sums of Hamiltonians, paraphrasing what the authors said about the role of process additivity in their theory. 

Finally, it is important to stress that a stochastic dynamics is not necessarily at odds with time-reversibility, since according to a BTQFT if $t\mapsto{Q_{t}}$ is a possible history of the universe, then also its time reverse $t\mapsto{Q_{-t}}$ is a possible path of this theory.\footnote{For details see \cite{durr_bell_type_2005}, Sections 6.1 and 6.2.}  
\vspace{2mm}

Having qualitatively introduced BTQFT, it seems plausible to claim that the structure of the theory is not artificial, as Nikoli\'c claims. However, the rest of the section is devoted to support this thesis with further arguments.

First of all, let us consider in the context of BTQFT the example of the unstable particle discussed in Sec. 2. Here we have a configuration composed by a single particle which follows a continuous trajectory described by the Bohmian law \eqref{guide} and only two cases are realizable: either the particle does not decay and the evolution of the particle remains described by \eqref{guide}, or the particle decays and a jump occurs, giving raise to a creation event (this solution is valid regardless of whether we consider the particle free to move in space or within an isolated box). This example shows how BTQFT explains in a much easier and more natural way phenomena of particle creation and annihilation with respect to Nikoli\'c's account: easier, since we have a clear dynamical evolution for the system which is always in a well-defined configuration, avoiding ambiguous descriptions implying the existence of three particles in physical space and avoiding the need of an interaction with a measurement device to determine its state, more natural since the sea of dead particles is completely avoided, returning to a clearer Bohmian ontology.\footnote{With this sentence my intention is \emph{not} to claim that a Bohmian theory must necessarily implement a particle ontology: Bohm himself in the appendix of his \cite{Bohm:1952ab} proposes a field ontology to extend his theory to electromagnetism. Furthermore, in literature exist many attempts to formulate a field ontology in the context of Bohmian QFTs, see Sec. 3 of \cite{Struyve:2010aa} for an overview.}

In this regard, another positive feature of this model is the simplicity with which it explains the experimental evidence, since that has been always considered an important meta-empirical virtue of physical theories. 

Secondly, it is appropriate to claim that the stochastic part of the dynamics resembles the mathematically well-defined processes of the wave function collapses in GRW theories, since in both cases these stochastic processes are \emph{spontaneous} in a precise sense: they are not caused or induced by external factors as measurements, observers, forces, etc. More precisely, in the GRW theory the evolution of the wave function is given by stochastic jump processes in Hilbert space which are responsible for the random collapses of the wave function. Between these random processes it evolves deterministically according to the Schr\"odinger equation. As in the case of the BTQFT, the GRW formalism provides the rates for these collapses.\footnote{For details on the similar structures of BM and GWR see \cite{Allori:2008aa} and \cite{Allori:2013aa}.}
Thus, there is a structural similarity between the processes associated with the inherent motion of the primitive ontologies of these theories. 

It is also important to underline that BTQFT is based on a mathematical framework which is widespread in applied sciences. The piecewise deterministic Markov process proposed by these authors to account for the dynamics of the particles is a standard method used to analyze the evolution of a given class of individuals and its evolution in time, which may well include variations in the number of its components. 

Furthermore, it is important to stress again that these authors derive the jump rates from the Schr\"odinger equation, which is part of the usual structure of quantum theories. 

Finally, it has also been shown that from suitable choices of jump rates it is possible to recover the standard BM (see \cite{Tumulka:2005} and \cite{Vink:1993}) as well as Nelson stochastic mechanics. 

All these facts, in turn, imply that it seems too strong to claim that the BTQFT relies on artificial structures. If the aim of the PO approach to quantum physics is to provide a methodology to construct rigorous theories which avoid the puzzles of the standard QM and QFT, then the BTQFT belongs properly to this family. All these theories must have what has been called a ``common structure'' in \cite{Allori:2008aa}, which includes the following conditions: 
\begin{enumerate}
   \item a clear PO specifying the distribution of matter in space;
   \item a state vector $\psi\in\mathcal{H}$ which evolves unitarily;
   \item $\psi$ governs the evolution of the PO by means of either deterministic or stochastic laws;
   \item the theory provides typical histories of the PO which are consistent with the quantum formalism,
\end{enumerate}
Clearly this structure is respected by the BTQFT. In sum, BM and BTQFT are simply different instantiations of a quantum theory without observers. Thus, it seems that Nikoli\'c's concern about the artificiality and naturalness of BTQFT is not problematic for someone willing to consider this theory as a serious extension of BM to QFT.

\section{Conclusion}

In this paper I have argued that Nikoli\'c's objections advanced against the Bell-type QFT in his \cite{Nikolic:2010aa} are well supported neither from a technical point of view, nor from an ontological one; as I have shown his theory implies a number of problematic aspects which allow one to discard this proposal.
Furthermore, it must be said that a deterministic characterization of the phenomena of particle creation and annihilation has been provided by \cite{Colin:2003aa} and \cite{Colin:2007aa}, rephrasing in Bohmian terms the Dirac Sea (DS) approach. The DS picture yields a more natural explanation of these events with respect to Nikoli\'c's theory, avoiding the problems arising with the derivation of the unwelcome sea of dead particles, which remarkably differs from the notion of Dirac sea. Moreover, while it is correct to consider Bell's proposal for a Bohmian QFT as a heuristic theory, the same judgement does not hold in the case of BTQFT. The latter theory avoids the difficulties of Bell's model, while successfully accounts for the phenomena of particle creation and annihilation, providing a more satisfactory characterization for them with respect to Nikoli\'c's theory. Finally, I have stressed not only that it is historically false to claim that the aim of BM is to restore determinism in physics, but also how inherent stochasticity can be perfectly compatible with a Bohmian theory, insofar as it respects the structure that a QTWO has to embed.
\vspace{5mm}

\textbf{Acknowledgemnts} 
I would like to thank Michael Esfeld, Anna Marmodoro, Davide Romano, Dustin Lazarovici, Mario Hubert, Olga Sarno and the anonymous referees for helpful comments on this paper. I am grateful to the Swiss National Science Foundation for financial support (Grant No. 105212-175971).
\clearpage

\bibliographystyle{apalike}
\bibliography{PhDthesis}
\end{document}